# ACOUSTICAL CLASSIFICATION OF DIFFERENT SPEECH ACTS USING NONLINEAR METHODS

Chirayata Bhattacharyya[1], Sourya Sengupta[2], Sayan Nag[2], Shankha Sanyal*[3,4], Archi Banerjee[3,4], Ranjan Sengupta[3] and Dipak Ghosh[3]

[1] Department of Electronics and Telecommunications Engineering, Jadavpur University
[2] Department of Electrical Engineering, Jadavpur University
[3] Sir C.V. Raman Centre for Physics and Music, Jadavpur University
[4] Department of Physics, Jadavpur University

*ssanyal.sanyal2@gmail.com

**ABSTRACT:**

*A recitation is a way of combining the words together so that they have a sense of rhythm and thus an emotional content is imbibed within. In this study we envisaged to answer these questions in a scientific manner taking into consideration 5 (five) well known Bengali recitations of different poets conveying a variety of moods ranging from joy to sorrow. The clips were recited as well as read (in the form of flat speech without any rhythm) by the same person to avoid any perceptual difference arising out of timbre variation. Next, the emotional content from the 5 recitations were standardized with the help of listening test conducted on a pool of 50 participants. The recitations as well as the speech were analyzed with the help of a latest non linear technique called Detrended Fluctuation Analysis (DFA) that gives a scaling exponent α, which is essentially the measure of long range correlations present in the signal. Similar pieces (the parts which have the exact lyrical content in speech as well as in the recital) were extracted from the complete signal and analyzed with the help of DFA technique. Our analysis shows that the scaling exponent for all parts of recitation were much higher in general as compared to their counterparts in speech. We have also established a critical value from our analysis, above which a mere speech may become a recitation. The case may be similar to the conventional phase transition, wherein the measurement of external condition at which the transformation occurs (generally temperature) is called phase transition. Further, we have also categorized the 5 recitations on the basis of their emotional content with the help of the same DFA technique. Analysis with a greater variety of recitations is being carried out to yield more interesting results.*

## INTRODUCTION

Audio signals, after the visual medium, are perhaps the most common phenomena to carry the essence of emotion breaking all kinds of barriers. Acoustic research over last few years has seen a great positive change with the advent of various neuro computational techniques. Numerous neuro cognitive techniques along with the novel non linear dynamics of chaos theory have made it possible to simulatanoeusly examine the analysis of acoustics and the emotional aspects they carry at brain [1]. Although musicsignals have provided a great number of tools of research in this aspect, the present study deals with the analysis of speech signals, more sepcifically when they are rendered with the essence of recitation. It can be looked as an alternative study of examining music, the meldoies kept apart from the text and the own flavours of proper emotions evoked by the meaning added to it.

From a physical point of view, such kinds of signals with proper rhythm are approximately periodic in macro and micro forms [2]. In this context, these signals then might have shown some kind of deterministic behaviour, but that is actually not the case. Earlier studies reflect that different parts of such rhythmical signals show distinguished behaviours, much different than others. In our study, we have tried to find change in the mathematical behaviour of signal parts according to the various emotions they may evoke.All these recitation signals actually show a complex behaviour in both macro and micro level which are very much chaotic, self organized, and particularly non linear[3]. Therefore the approach of study should not have been fruitful if we choose some deterministic or linear ways to venture through. Hence, non linear dynamical modelling of source comes into picture.

A number of studies provide support in the relevance of non-deterministic/chaotic approaches in understanding the speech signals [4]. Fractal analysis of the signals reveals the complex geometry embedded in signal. Fractal analysis of audio signals was first performed by Voss and Clarke, who came up with the analysis of amplitude spectra of audio signlas to find out some characteristic frequency fc, which seperates the white noise present in the signal at frequencis mush lower than fc [5].

Music or Recitation signals are practically quantitative record of variations of some particular qualities over a period of time. One way of analyzing it is to examine the geometric features to categorize the data [6]. The fractal signal is generally of two types, namely, multi-fractal and mono-fractal [7]. Fractal analysis is used to find the Hurst-exponent, singularity exponent and singularity dimension and therefore to find the multi-fractal spectrum. A number of studies have used Detrended Fluctuation Analysis (DFA) of EEG signals elicited by a variety of stimuli because of its robustness to non-stationary. Detrending involves the isolation of the low frequency variation (i.e. trend) and to decompose the residual signal into a seasonal (or cyclic) and a random walk type variation (i.e. white noise or high frequency noise) [8]. There are also a few reports that have used this technique to study the scaling behavior of the fluctuations in the music signal, as well as in the detection of arousal based effects in music induced EEG signals [9].

In our study, we have tried to correlate how the Hurst exponent varies when the mode of reading a text changes from normal speech to recitation, i.e. we add an inherent rhythmic pattern to the speech signals. The emotional aspects of the texts involved has been actually mapped with their auto correlative measure in terms of the Hurst Exponent parameter [10].

Our study also reflects the need of evoking emotions related with the texts properly. Questions may arise, does choosing a set of five regional poems make the study a local one? The answer from our part is successfully, NO. The language is not at all a barrier when we talk about emotion. A text recited with proper emotional aspects can arise the same feeling of anger, love, pain or serenity or whatever it may be associated with, inside the mind of an individual irrespective of the language difference between them. The meanings may be left not understood, but the emotions can be carried in the same way.

The results reveal interesting details of the mapping between emotion and mathematics clearly.

## EXPERIMENTAL DETAILS

Five widely known Bengali poems of different genres has been taken as subjects of experiment. Each of them expressing a different mode can surge different emotional aspects when read in proper manner. These different poems have been listed below as per emotions they can surge, varying over the respective genres.

**Table 1:** Different poems taken as subjects for our analysis.

| Poem | Poet | Genre |
|---|---|---|
| Khuror Kwol | Sukumar Ray | Fun (Nonsense Laughter) |
| Lukochuri | Rabindranath Tagore | Bliss |
| Jol Haoar Lekha | Joy Goswami | Romance (Less Rhythmic, slower) |
| Premikjoner Chithi | Srijato Bandyopadhyay | Romance (More Rhythmic, Faster) |
| Bujhbe Sedin Bujhbe | Kaji Najrul Islam | Pain |

The signals are digitized at the rate of 22050 samples/sec 16 bit format. Recital or free reading of each poem has been kept within a time duration of more or less 2 minutes and similar phases (which were about 35-40 seconds duration) has been extracted then after for the ease of analysis. The fractal analysis of different segments has been carried out separately to get the necessary Hurst exponent measures.

## METHODOLOGY
**Detrended Fluctuation Analysis (DFA):**
DFA is an interesting method for scaling the long-term autocorrelation of non-stationary signals. It quantifies the complexity of signals using the fractal property. DFA was first proposed by Peng et al. in 1995. This method is a modified root mean square method for the random walk. Mean square distance of the signal from the local trend line is analyzed as a function of scale parameter. There is usually power-law dependence and interesting parameter is the exponent. In many cases the DFA scaling exponent can be used to discriminate healthy and pathological data.

**DFA Algorithm:**
The procedures to compute DFA of a time series $[x_1, x_2,...,x_N]$ are as follows [11].
(1) First integrate x into a new series y = [y(1),...,y(N)], where $y(k)=\sum_{i=1}^{k}(x_i - \bar{x})$ and $\bar{x}$ is the average of $x_1, x_2,...,x_N$.
(2) The integrated series is then sliced into boxes of equal length n. In each box of length n, a least-squares line is fit to the data, representing the trend in that box. The coordinates of the straight line segments are denoted by $y_n(k)$.
(3) The root-mean-square fluctuation of the integrated series is calculated by $F(n)= \sqrt{(1/N)\sum_{k=1}^{N}[y(k)-y_n(k)]}$, where the part $[y(k)- y_n(k)]$ is called detrending.
(4) The relationship between the detrended series and interval lengths can be expressed as $F(n) \propto n^\alpha$ where α is expressed as the slope of a double logarithmic plot log [F(n)] versus log(n). The parameter α (scaling exponent, autocorrelation exponent, self-similarity parameter) represents the autocorrelation properties of the signal.

**Scaling exponent**
The parameter α (scaling exponent, autocorrelation exponent, self-similarity parameter) represents the autocorrelation properties of the signal
1. α < 0.5 anti-correlated signal
2. α = 0.5 uncorrelated signal (white noise)
3. α > 0.5 positive autocorrelation in the signal
4. α = 1 1/f noise
5. α = 1.5 Brownian noise or random walk

Using scaling exponent α one should be able to completely describe the significant autocorrelation properties of the bio-medical signals. Often computed separately exponent for low and high n can describe short-range scaling exponent (or fast parameter) $α_1$ and long-range scaling exponent (or slow parameter) $α_2$ for time scales [12].

## RESULTS AND DISCUSSION

The DFA exponent corresponding to each portion of the recitation clips chosen for our analysis have been reported in **Table 2.** Each clip was divided into three segments of 30 seconds each and the DFA scaling exponent corresponding to each segment was calculated to get an estimate of the long range temporal correlations present in each of the segment. The DFA exponent has been used as a quantitative parameter which distinguishes the time series data originating from the two modes of speech and recitation.

**Table 2:** DFA exponent corresponding to normal reading and recitation of the five clips

|  | Poem | Genre of Emotion | Segment No. | Hurst Exponent (Free Reading) | Hurst Exponent (Recitation) |
|---|---|---|---|---|---|
| Clip 1 | Khuror Kwol | Fun (Nonsense Laughter) | #1 | 0.314 | 0.216 |
|  |  |  | #2 | 0.324 | 0.236 |
|  |  |  | #3 | 0.293 | 0.206 |
| Clip 2 | Lukochuri | Happy | #1 | 0.303 | 0.221 |
|  |  |  | #2 | 0.322 | 0.236 |
|  |  |  | #3 | 0.316 | 0.237 |
| Clip 3 | Jol Haoar Lekha | Romance (Slower) | #1 | 0.353 | 0.254 |
|  |  |  | #2 | 0.302 | 0.270 |
|  |  |  | #3 | 0.326 | 0.250 |
| Clip 4 | Premikjoner Chithi | Romance (Fast Rhythm) | #1 | 0.379 | 0.290 |
|  |  |  | #2 | 0.372 | 0.284 |
|  |  |  | #3 | 0.322 | 0.293 |
| Clip 5 | Bujhbe Sedin Bujhbe | Sorrow | #1 | 0.368 | 0.421 |
|  |  |  | #2 | 0.376 | 0.429 |
|  |  |  | #3 | 0.380 | 0.403 |

The averaged values corresponding to each clip of normal reading and recitation have been plotted in **Fig. 1** and the corresponding error bars represent the SD values generated from each clip.

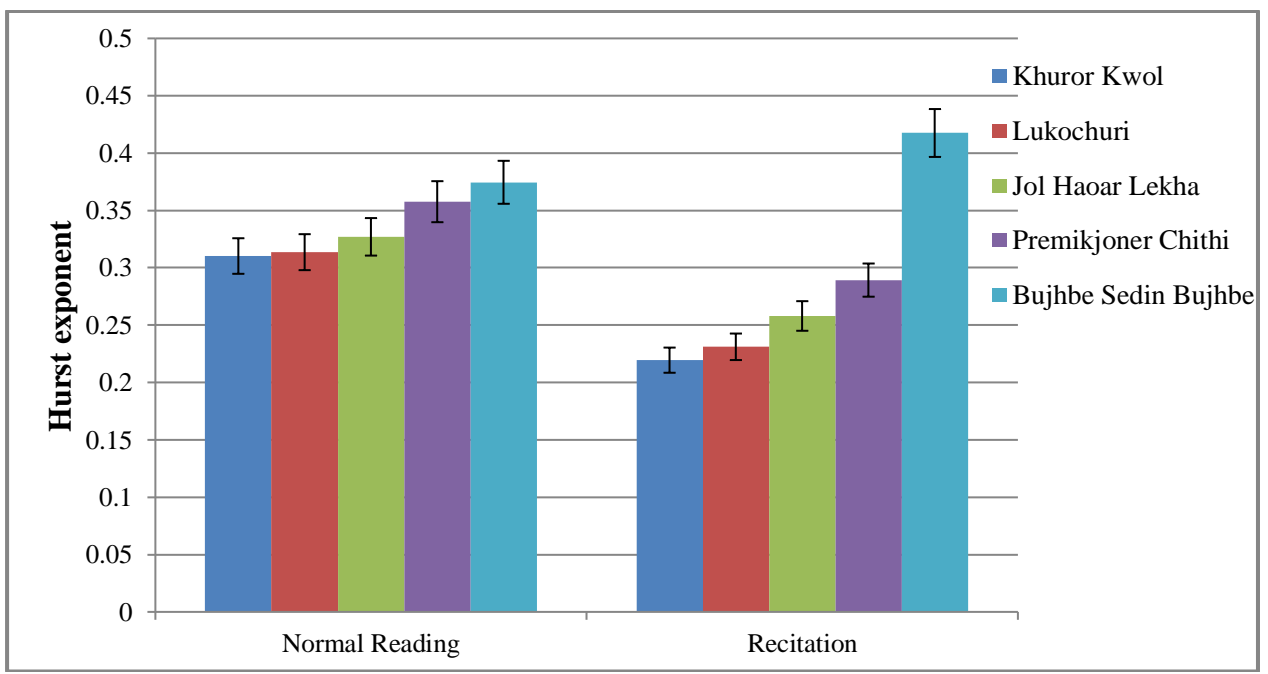

**Fig. 1:** Averaged DFA exponent for each clip corresponding to normal reading and recitation

In general, it is seen that for any mode of speech, the DFA exponent is on the lower side, i.e. in most cases, it is below 0.5, which signifies a lower correlation in the time series data of speech mode. A general glance at the figures show that in general, the Hurst exponent corresponding to normal reading is mostly on the higher side compared to their recitation counterpart. An exception is observed for the last clip "*Premikjoner Chithi*", whereby the Hurst exponent for normal reading is lower than the Hurst exponent for recitation case. To get an estimate of the difference between the two phases of reading and recitation, the following **Fig. 2** have been plotted:

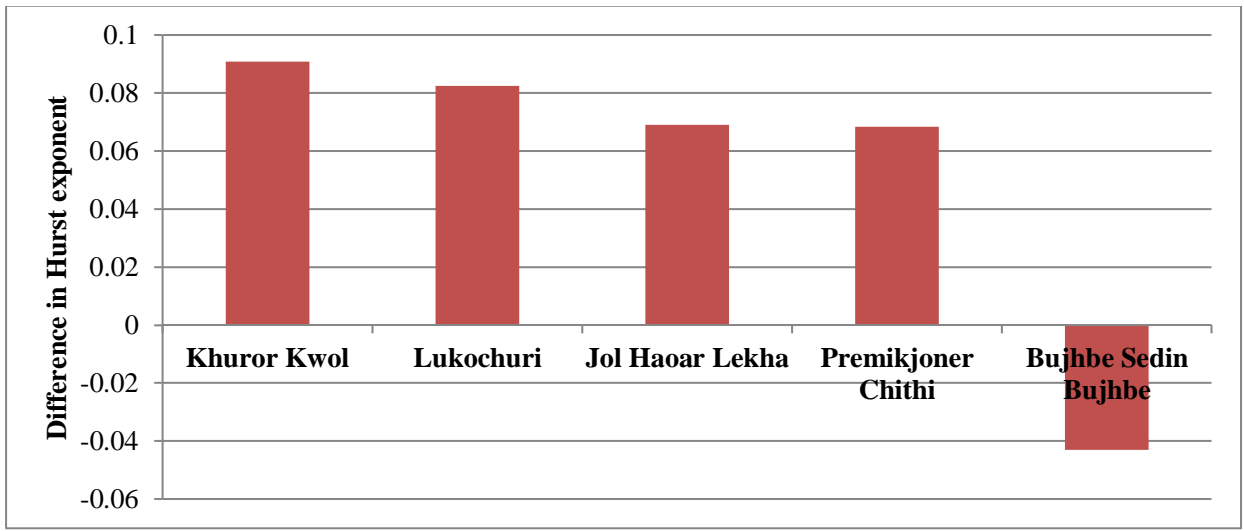

**Fig. 2:** Difference in Hurst exponents between reading and recitation phase

It is seen from the figure, that the first poem registers the maximum change during the change in phase from normal reading to recitation. This is followed by the 2$^{nd}$ poem "*Lukochuri*", and the rest. The 3$^{rd}$ and 4$^{th}$ poem have almost similar amount of change but in the lower side. Only the last poem, "*Bujhbe Sedin Bujhbe*" provides a decrease in the value of Hurst exponent.

Furthermore, a cue for emotion detection is also revealed from this study. The pieces have been chosen in such a way that each of them belong to a different genre of emotion. As is seen from the **Fig. 1**, the Hurst exponents belonging to different emotional categories are significantly different. For the poems belonging to positive emotional valence, we see the DFA exponents are on the lower side. As the emotional intensity changes from happy to romantic, the Hurst exponent increases correspondingly. In this regard, it is also interesting to note that the two romantic clips (viz. Clips 4 and 5) have almost similar Hurst exponents, though the lyrical content of the two poems are completely different from one another. For sad emotion, we find the value of Hurst exponent to be the maximum amongst all the clips for Clip 5. In this way, without the analysis of any prosodic features, we provide a quantitative robust algorithm with the help of which the speech and recitation part of a clip can be segregated. Also, the emotional content of a recitation can be identified.

## CONCLUSION:

In this work, for the first time an attempt is being made to develop a new robust algorithm with which normal speech can be differentiated from recitation just like phase transformation in classical physics. The following are the main conclusions from the study:

1. The DFA exponent for any mode of speech is found to be lower than that compared to music [13] as is seen from earlier works. This signifies the correlations present in the time series data of speech related signals is on the lower side as compared to music or any other form of acoustic signals.
2. The Hurst exponent of normal speech is always on the higher side as compared to the recitation of the same lyrical content. This implies that the addition of melodic content decreases the long range temporal correlations in the acoustic signal.

3. We have identified a threshold value of 0.3 for Hurst exponent, beyond which the transformation from speech to recitation occurs.
4. A hint in the direction of emotion quantification is also obtained from this work, whereby we see that the Hurst exponent decreases as the emotional content of the recitation changes from positive to negative.

For the first time, a quantification study attempts to categorize the emotional content of a speech data and also to segregate two formats of reading. Further works are being pursued with a greater number of samples taken for analysis to have a more conclusive outcome. This is a pilot study in that direction.